\documentclass{emulateapj} 
\usepackage{natbib}
\usepackage{epsfig}
\usepackage{graphicx} 
\usepackage{subfigure}
\usepackage{float}
\usepackage{amsmath}
\usepackage{color}
\usepackage{amssymb}
\usepackage{amsfonts}
\usepackage{units}
\usepackage{mathtools}
\usepackage{bm}
\usepackage[colorlinks,linkcolor=blue,anchorcolor=green,citecolor=blue]{hyperref}
\def\be{\begin{eqnarray}} 
\def\ee{\end{eqnarray}}

\shorttitle{Synchrotron radiation with pitch-angle distribution}
\shortauthors{Yang \& Zhang} 
\begin{document} 

\title{Synchrotron radiation from electrons with a pitch-angle distribution}

\author{Yuan-Pei Yang\altaffilmark{1,2,3} and Bing Zhang\altaffilmark{4,2}}

\affil{$^1$Kavli Institute for Astronomy and Astrophysics, Peking University, Beijing 100871, China; yypspore@gmail.com;\\
$^2$ National Astronomical Observatories, Chinese Academy of Sciences, Beijing 100012, China\\
$^3$ KIAA-CAS Fellow\\ 
$^4$ Department of Physics and Astronomy, University of Nevada, Las Vegas, NV 89154, USA; zhang@physics.unlv.edu
}

\begin{abstract}
In most astrophysical processes involving synchrotron radiation, the pitch-angle distribution of the electrons is assumed to be isotropic. However, if electrons are accelerated anisotropically, e.g, in a relativistic shock wave with an ordered magnetic field or in magnetic reconnection regions, the electron pitch angles might be anisotropic. In this work, we study synchrotron radiation from electrons with a pitch-angle distribution with respect to a large-scale uniform magnetic field. Assuming that the pitch-angle distribution is normal with a scatter of $\sigma_p$ and that the viewing direction is where the pitch-angle direction peaks, we find that for electrons with a Lorentz factor $\gamma$, the observed flux satisfies $F_\nu\propto\nu^{2/3}$ for $\nu\ll\nu_{\rm cr}$ ($\nu_{\rm cr}$ is the critical frequency of synchrotron),  if $\sigma_p\lesssim1/\gamma$ is satisfied. On the other hand, if $\sigma_p\gg1/\gamma$, the spectrum below $\nu_{\rm cr}$ is a broken power law with a break frequency $\nu_{\rm br}\sim2\nu_{\rm cr}/\sigma_p^3\gamma^3$, e.g., $F_\nu\propto\nu^{2/3}$ for $\nu\ll\nu_{\rm br}$ and $F_\nu\propto\nu^{1/3}$ for $\nu_{\rm br}\ll\nu\ll\nu_{\rm cr}$. Thus the ultimate synchrotron line of death is $F_\nu\propto\nu^{2/3}$. We discuss the application of this theory to blazars and gamma-ray bursts (GRBs).
\end{abstract}


\section{Introduction}

Synchrotron radiation is one of non-thermal radiation mechanisms that plays an important role in many astrophysical sources. The classical synchrotron theory states that the radiation power from an electron is proportional to $\nu^{1/3}$ below a characteristic frequency \citep[e.g.][]{ryb79,jac98}. This feature seems supported by some observations from astrophysical sources. However, we should note that the $\nu^{1/3}$ spectrum from an electron corresponds to the angle-integrated radiation rather than radiation per solid angle along the line of sight. The latter has a spectrum $\nu^{2/3}$ at low frequencies \citep[e.g.][]{ryb79,jac98}. In most astrophysical processes, since an underlying assumption is that the pitch-angle distribution of electrons is isotropic, the classical $\nu^{1/3}$ spectrum is taken as the default. Such a spectrum is regarded as a ``line of death'' for synchrotron radiation \citep[e.g.][]{pre98}, and deviation of this spectrum has been regarded evidence against synchrotron as the radiation mechanism.

On the other hand, the momentum distribution of electrons might appear significantly anisotropic if there exists ordered magnetic fields in the emission region. When electrons move helically in a large-scale magnetic field, the pitch-angle distribution would be anisotropic\footnote{The pitch angle here is defined as the angle between the the electron momentum and the magnetic field line.}, which would cause the deviation from the classical $\nu^{1/3}$ spectrum. For the most extreme case that the angle distribution is a Delta function, the observed spectrum would satisfy $F_\nu\propto\nu^{2/3}$ at low frequencies.

In this work, we consider synchrotron radiation from electrons with a pitch-angle distribution with respect to a large-scale magnetic field.
The theoretical framework is laid out in Section 2. The synchrotron spectra under different conditions are presented in Section 3. Some astrophysical applications are discussed in Section 4, and the results are summarized in Section 5.

\section{Synchrotron radiation with a pitch angle distribution}\label{sec1}

\begin{figure}[]
\centering
\includegraphics[angle=0,scale=0.35]{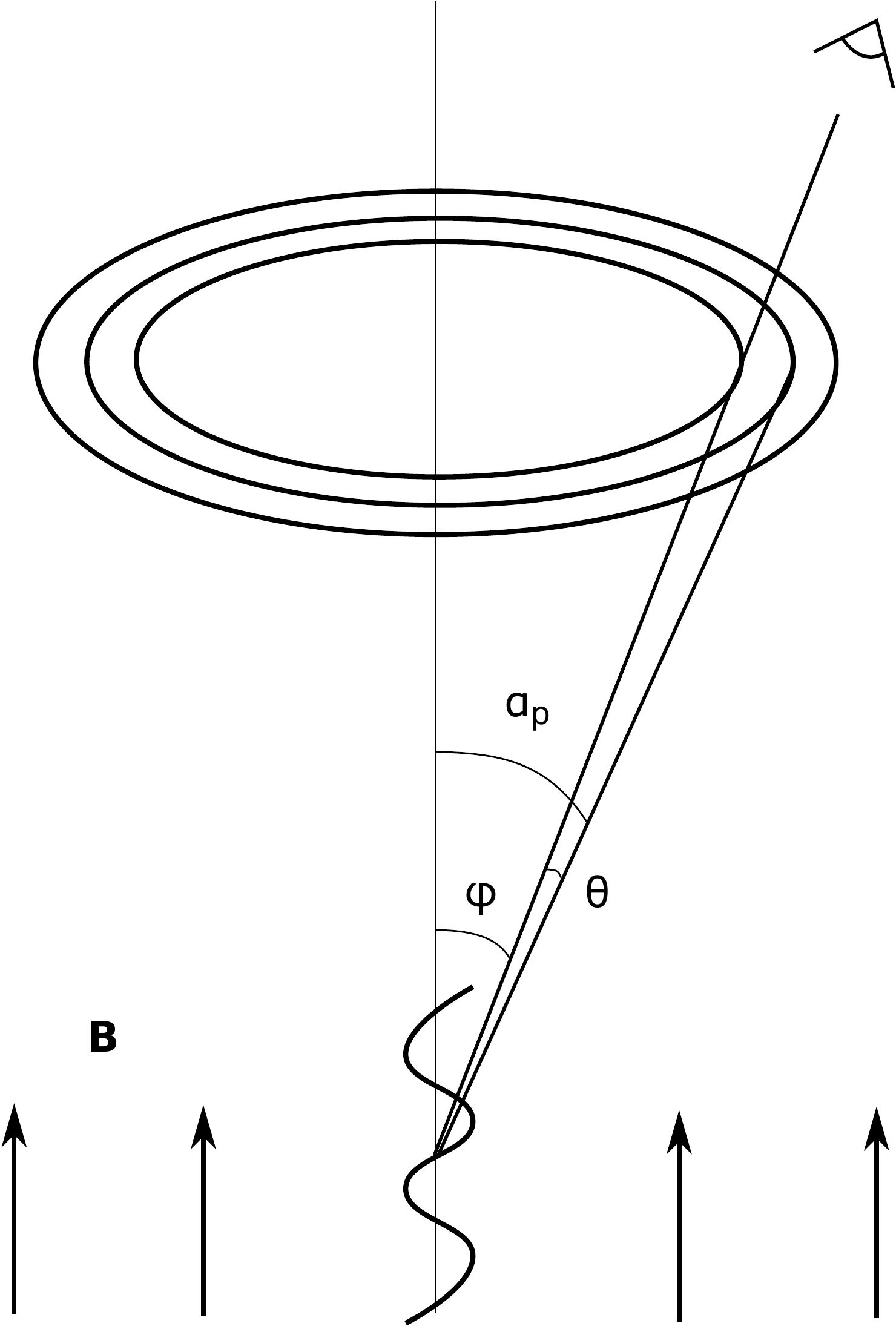} 
\caption{Synchrotron radiation from an electron with a pitch angle $\alpha_p$. The viewing angle is $\varphi$. For a given frequency $\nu$, radiation is confined to the circles with $\theta\sim\pm\theta_c(\nu)$.}\label{cart1}
\end{figure}

We assume that the magnetic field is large-scale uniform.
The motion of an electron in a uniform magnetic field is helical with a pitch angle $\alpha_p$, as shown in Figure \ref{cart1}. The motion can be decomposed into the components along the field line and in the plane normal to the field. The gyration frequency is $\omega_B=eB/\gamma m_ec$, where $B$ is the magnetic field strength.
The radius of curvature of the electron path is 
\be
\rho=\frac{1}{\sin\alpha_p}\frac{\gamma m_ec^2}{eB}.
\ee
Here we have assumed that the velocity component in a plane normal to the field is relativistic. Note that this differs by a factor $\sin\alpha_p$ from the radius of the projected circle in the plane normal to the field.

We consider an instantaneously-circular motion with a curvature radius $\rho$ and define the angle between the line of sight and the instantaneous-trajectory plane as $\theta$ (see Figure \ref{cart1}).  
The received power per unit frequency interval per unit solid angle is given by \citep[e.g.][]{jac98}
\be
\frac{dW}{d\omega d\Omega}&=&\left(\frac{\omega_B}{2\pi}\frac{1}{\sin^2\alpha_p}\right)\frac{e^2}{3\pi^2c}\left(\frac{\omega\rho}{c}\right)^2\nonumber\\
&\times&\left(\frac{1}{\gamma^2}+\theta^2\right)^2\left[K_{2/3}^2(\xi)+\frac{\theta^2}{(1/\gamma^2)+\theta^2}K_{1/3}^2(\xi)\right],\nonumber\\\label{CRspec}
\ee
where $\omega_B/2\pi$ corresponds to the reciprocal value of the gyration period, the factor $1/\sin^2 \alpha_p$ corresponds to the correction between received power and emitted power \citep[e.g.][]{ryb79}, and the parameter $\xi$ in the modified Bessel function $K_{\nu}(\xi)$ is defined by $\xi=(\omega\rho/3c)\left(1/\gamma^2+\theta^2\right)^{3/2}$ \citep{jac98}. 
The $K_{2/3}(\xi)$ term corresponds to the polarized component in the trajectory plane, and the $K_{1/3}(\xi)$ term corresponds to the polarized component that is perpendicular to the line of sight and the $K_{2/3}(\xi)$ component. Numerically, the radiation is dominated by the first term.

As shown in the above equation, the observed spectrum depends on the observational direction. At $\theta=0$, the radiation power reaches the maximum, and one has \citep[e.g.][]{jac98}
\be
\left.\frac{dW}{d\omega d\Omega}\right|_{\theta=0}\simeq
\frac{\gamma e^3B}{m_ec^2\sin^2\alpha_p}
\begin{dcases}
\frac{3\Gamma^2(2/3)}{2^{7/3}\pi^3}\left(\frac{\omega}{\omega_{\rm cr}}\right)^{2/3}, &\omega\ll\omega_{\rm cr}, \\
\frac{3}{8\pi^2}\frac{\omega}{\omega_{\rm cr}}e^{-\omega/\omega_{\rm cr}}, &\omega\gg\omega_{\rm cr}.
\end{dcases} \nonumber\\\label{scr}
\ee
where $\omega_{\rm cr}$ is the critical frequency of the curvature radiation, i.e.
\be
\omega_{\rm cr}=\frac{3}{2}\gamma^3\left(\frac{c}{\rho}\right). \label{omegac}
\ee
Therefore, one has $dW/d\omega d\Omega\propto\omega^{2/3}$ for $\omega\ll\omega_{\rm cr}$. For the case with $\theta\neq0$, as $\theta$ increases the cut-off frequency of the spectrum shifts to lower frequencies, but the spectral index remains $2/3$. 

The spectrum of the received power can be found by integrating Eq.(\ref{CRspec}) over angle\footnote{Note that Eq.(\ref{syn}) corresponds to the received power, which is a factor of $1/\sin^2\alpha_p$ with respect to the emitted power \citep{ryb79}.} \citep{wes59}, i.e.
\be
\frac{dW}{d\omega}=\frac{\sqrt{3}e^3B}{2\pi m_ec^2\sin\alpha_p}\frac{\omega}{\omega_{\rm cr}}\int_{\omega/\omega_{\rm cr}}^\infty K_{5/3}(x)dx.\label{syn}
\ee
The above equation can give the classical angle-integrated spectrum of synchrotron radiation, i.e., $dW/d\omega\propto\omega^{1/3}$ for $\omega\ll\omega_{\rm cr}$, which applies to most astrophysical sources that invoke random magnetic fields or random pitch angles \citep[e.g.][]{yan17}.

We now investigate how the pitch-angle $\alpha_p$ distribution of the electrons affects the synchrotron spectrum. First, we assume that the electrons have a pitch-angle distribution of the form
\be
N_e(\alpha_p,\gamma)d\alpha_p d\gamma&=&\frac{1}{\sqrt{2\pi\sigma_p^2}}\exp\left(-\frac{(\sin\alpha_p-\sin\alpha_{p,0})^2}{2\sigma_p^2}\right)\nonumber\\
&\times&f_e(\gamma)d\alpha_p d\gamma,
\ee
where $\sin\alpha_{p,0}$ corresponds to the mean value of $\sin\alpha_p$, $\sigma_p$ is the corresponding standard deviation, and $f_e(\gamma)$ denotes the energy distribution of electrons. For mono-energy electrons, we define
\be
f_e(\gamma)=N_{e,0}\delta(\gamma-\gamma_0),
\ee 
where $N_{e,0}$ corresponds to the total number of electrons.
For a power-law distribution, we define
\be
f_e(\gamma)=\frac{N_{e,0}(p-1)}{\gamma_1}\left(\frac{\gamma}{\gamma_1}\right)^{-p}\label{power}
\ee
with $\gamma_1<\gamma<\gamma_2$.
 
One is allowed to take the element of solid angle to be $d\Omega=2\pi\sin\alpha_p d\alpha_p$.
Assuming that the view direction is $\varphi$, one has $\theta=\alpha_p-\varphi$, as shown in Figure \ref{cart1}. The observed spectrum at $\varphi$ would be given by
\be
F_\omega(\varphi,\omega)&=&\frac{2\pi}{D^2}\int_{-\varphi}^{\pi/2-\varphi}\int_{\gamma_1}^{\gamma_2} \frac{dW(\theta,\omega,\gamma)}{d\omega d\Omega}\nonumber\\
&\times&N_e(\varphi+\theta,\gamma)\sin(\varphi+\theta)d\gamma d\theta. \label{specm1}
\ee
 
In some astrophysical sources, e.g., GRBs or blazars, the emission region is in a relativistic shell/blob. The specific flux in the observer frame is given by \citep[e.g.][]{der09}
\be
F_\omega(\varphi,\omega)&=&\frac{2\pi\delta_D^3}{D^2}\int_{-\varphi}^{\pi/2-\varphi}\int_{\gamma_1}^{\gamma_2} \frac{dW(\theta,\omega/\delta_D,\gamma)}{d\omega d\Omega}\nonumber\\
&\times&N_e(\varphi+\theta,\gamma)\sin(\varphi+\theta)d\gamma d\theta,\label{specm2}
\ee
where $\delta_D$ is the Doppler factor.
Note that $\theta$, $\varphi$, $\gamma$, $\alpha_p$, $\Omega$, $W$, $N_e$ and $B$ are in the comoving frame.
The flux per frequency interval $d\nu$ is $F_\nu(\nu)=2\pi F_{\omega}(\omega)$ with $\nu=\omega/2\pi$.

\section{Results}

\begin{figure}[]
\centering
\includegraphics[angle=0,scale=0.4]{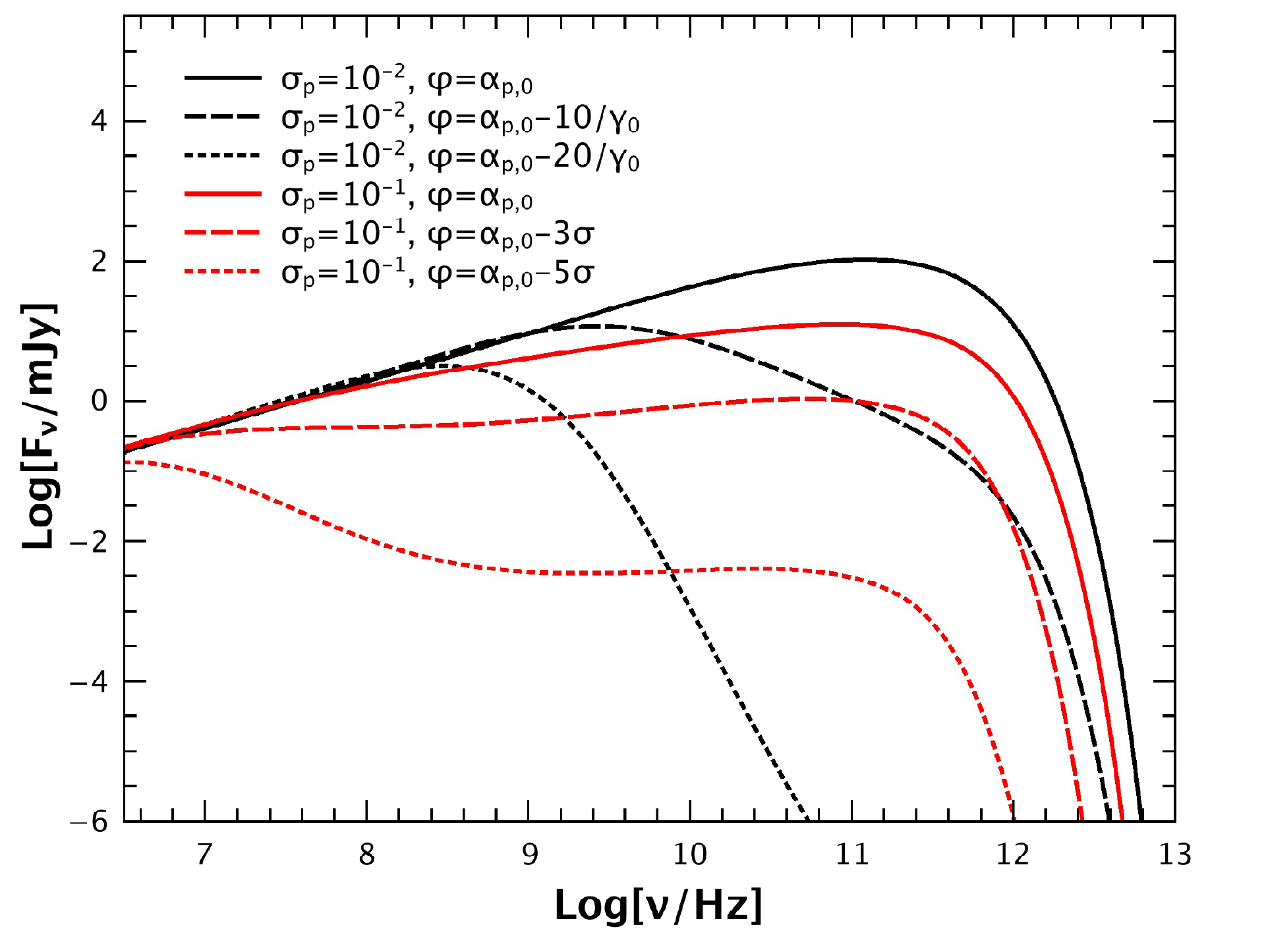}
\caption{Synchrotron spectra with different viewing directions. Following parameters are adopted: $\gamma_0=10^2$, $\alpha_{p,0}=\pi/4$, $B=1~\unit{G}$, $\delta_D=10$, $N_{e,0}=10^{48}$, and $D=1~\unit{Gpc}$.}\label{fig1}
\end{figure}

We first consider how the observed spectrum depends on the viewing direction $\varphi$. We assume that the electron distribution is mono-energetic, i.e., $\gamma\sim\gamma_0$. For electrons with a normal distribution of the pitch angles, the observed peak flux is obviously at the maximum when $\varphi=\alpha_{p,0}$. 
If $\sigma_p\lesssim1/\gamma$, once 
$\varphi\gtrsim\alpha_{p,0}-1/\gamma$ is reached, the observed spectrum is significantly suppressed, as shown in the black curves in Figure \ref{fig1}. The reason is that for $\sigma_p\lesssim1/\gamma$ the radiation is beamed in a narrow cone of the size of $\sim1/\gamma$. The radiation significantly decreases outside the $\sim1/\gamma$ cone. On the other hand, if $\sigma_p\gg1/\gamma$, the observed spectrum is not significantly affected by the viewing direction, as long as the viewing direction is within the pith-angle scatter $\sigma_p$. The observed spectrum is suppressed only when the viewing direction is outside the pitch-angle scatter, i.e., $\varphi\gtrsim\alpha_{p,0}-\sigma_p$, as shown as the red curves in Figure \ref{fig1}. 

We next focus on the case with $\varphi=\alpha_{p,0}$, which corresponds to the line of sight along the pitch-angle direction where the electron distribution reaches the maximum. This is a relevant geometry since it is the configuration where the observed emission is the brightest. We define the spectral index as
\be 
q\equiv \frac{d\ln F_\nu}{d\ln\nu}.
\ee
As shown in Figure \ref{fig2}, for mono-energetic electrons with $\gamma\sim\gamma_0$, the spectral index $q$ mainly depends on $\sigma_p$. If $\sigma_p\ll1/\gamma$, the spectrum would appear independent of $\sigma_p$, and one has $F_\nu\propto\nu^{2/3}$ for $\nu\ll\nu_{\rm cr}$. If $\sigma_p\gtrsim1/\gamma$, the spectrum would become softer, and there is a break frequency $\nu_{\rm br}$ below $\nu_{\rm cr}$ at
\be
\nu_{\rm br}\sim\frac{2\nu_{\rm cr}}{\sigma_p^3\gamma^3}, \label{break}
\ee
where $\nu_{\rm cr}=\omega_{\rm cr}/2\pi=3\gamma^2 eB\delta_D\sin\alpha_{p,0}/4\pi m_ec$ is the critical frequency of synchrotron radiation in the observed frame. 
In this case, one has $F_{\nu}\propto\nu^{2/3}$ if $\nu\ll\nu_{\rm br}$ and $F_{\nu}\propto\nu^{1/3}$ if $\nu_{\rm br}\ll\nu\ll\nu_{\rm cr}$. 

\begin{figure}[]
\centering
\includegraphics[angle=0,scale=0.3]{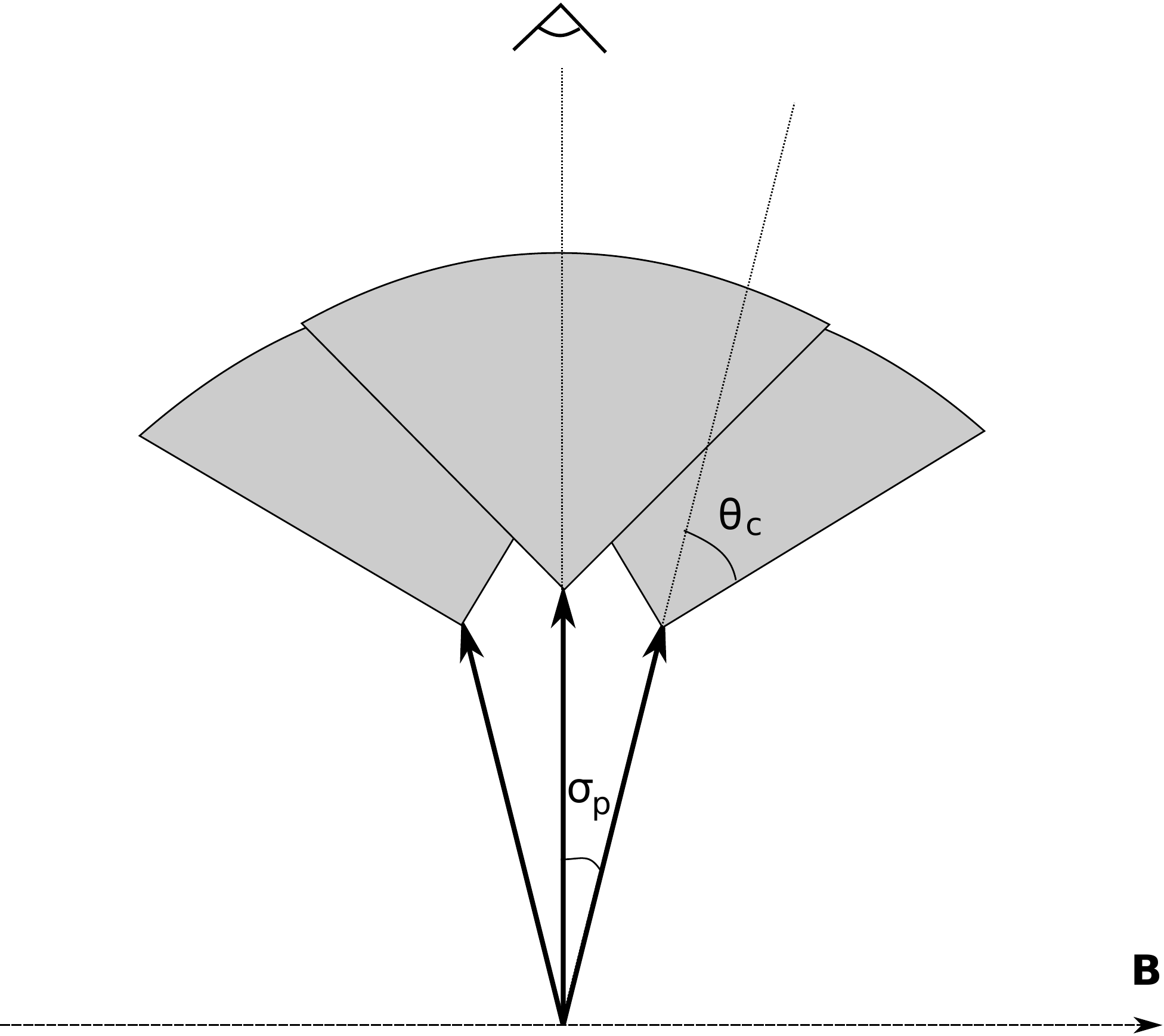}
\includegraphics[angle=0,scale=0.3]{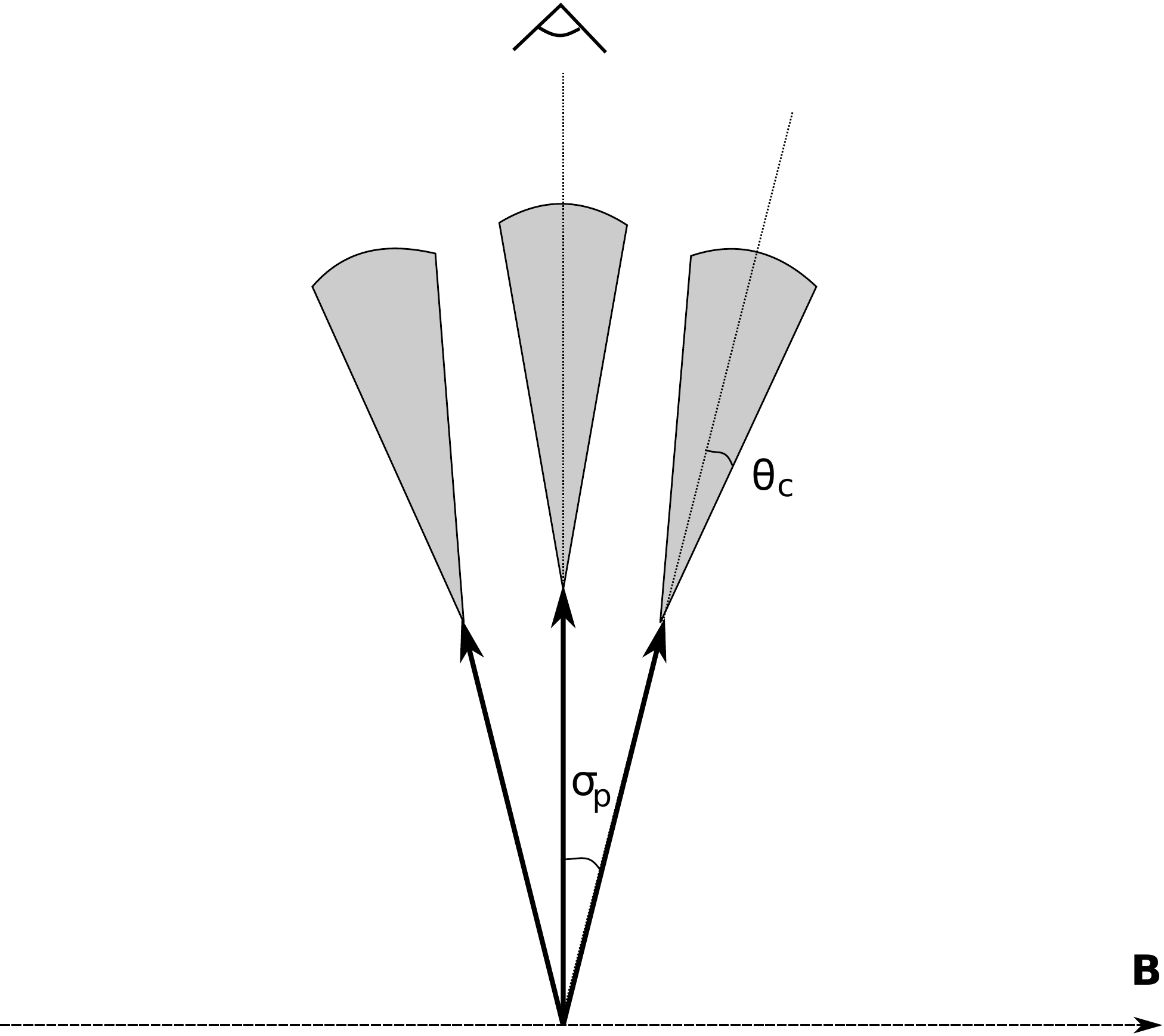} 
\caption{Synchrotron radiation from electrons with different pitch angles. Thick solid arrows denote the directions of electron momenta. Grey areas denote the synchrotron radiation beams of the electrons. The top panel corresponds to the electron beam being narrower than the radiation beam, i.e., $\sigma_p\ll\theta_c(\nu)$, and the bottom panel corresponds to the electron beam being wider than the radiation beam, i.e., $\sigma_p\gg\theta_c(\nu)$.  $\varphi=\alpha_{p,0}=\pi/2$ is adopted.}\label{cart2}
\end{figure}

\begin{figure}
\centering
\includegraphics[angle=0,scale=0.4]{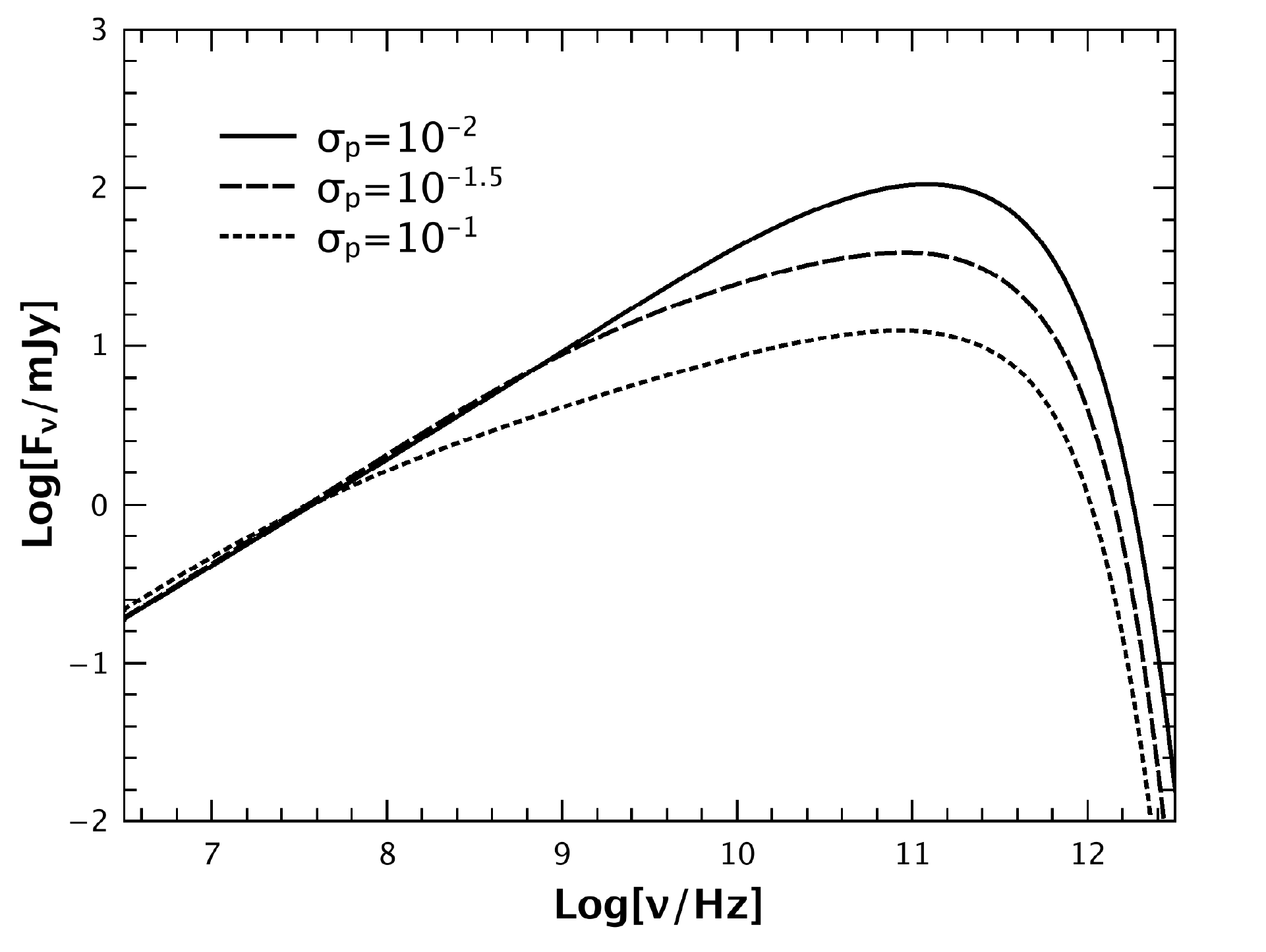}
\includegraphics[angle=0,scale=0.4]{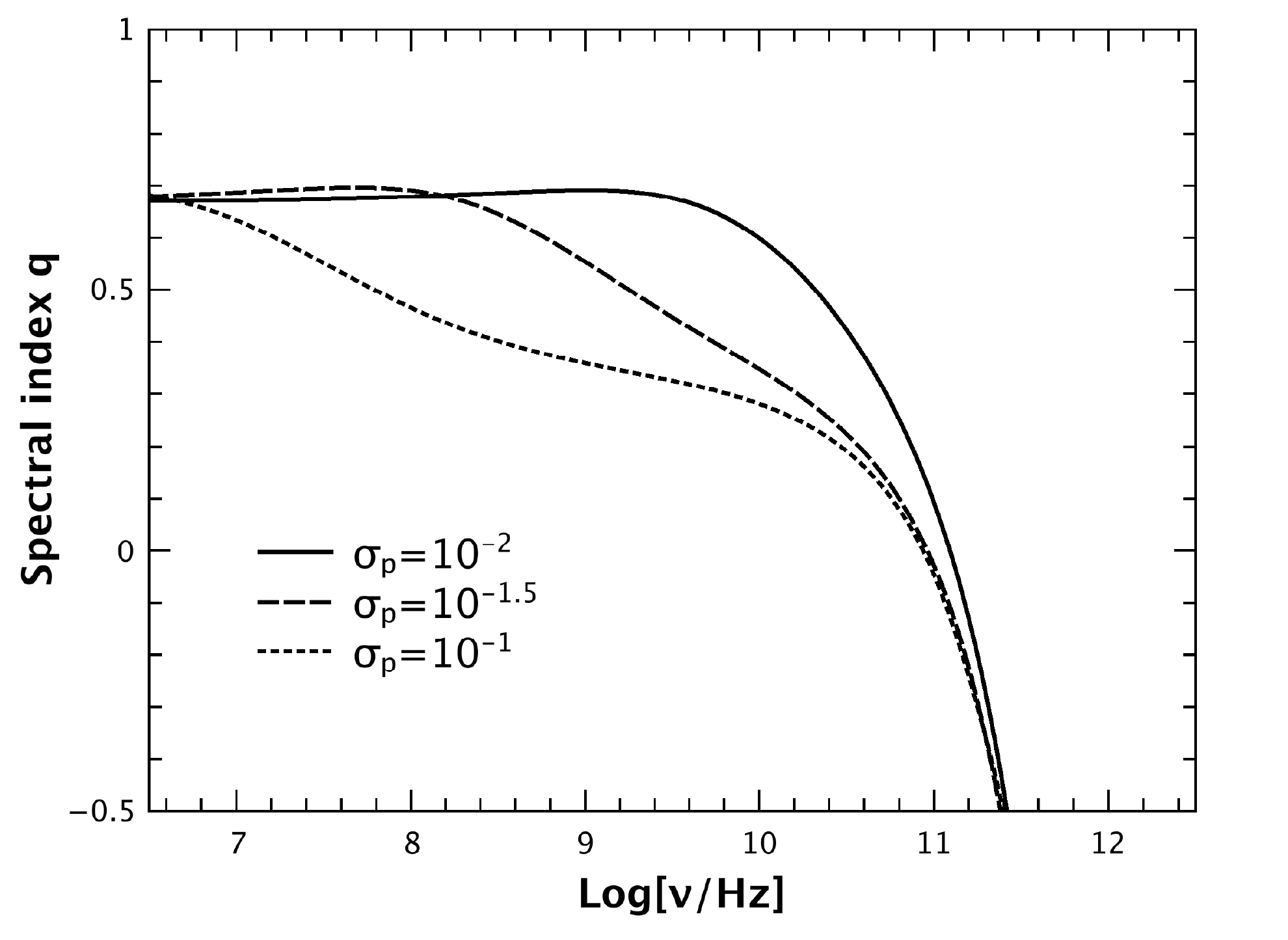}
\caption{Top panel: synchrotron spectra from mono-energetic electrons. Bottom panel: the relation between spectral index and frequency. Solid, dashed and dotted curves denote $\sigma_p=10^{-2}, \ 10^{-1.5}, \ 10^{-1}$, respectively. Following parameters are adopted: $\gamma_0=10^2$, $\alpha_{p,0}=\pi/4$, $B=1~\unit{G}$, $\delta_D=10$, $N_{e,0}=10^{48}$, and $D=1~\unit{Gpc}$.}\label{fig2}
\end{figure}

The reason for a broken pow-law spectrum is as follows: for synchrotron/curvature radiation from a single electron, the typical frequency-dependent spread angle is given by \citep[e.g.][]{jac98,yan17}
\be
\theta_c(\nu)\sim\frac{1}{\gamma}\left(\frac{2\nu_{\rm cr}}{\nu}\right)^{1/3}
\ee
for $\nu\ll\nu_{\rm cr}$. For a viewing direction within $\theta_c(\omega)$, the radiation angle power could be treated as the same. Once the viewing direction is outside $\theta_c(\nu)$, the radiation angular power would significantly decrease.
If $\nu\ll\nu_{\rm br}$, one has $\sigma_p\ll\theta_c(\nu)$, which means that the electron beam is much narrower than the radiation beam of a single electron.
Along the line of sight, for the electrons with different pitch angles, the radiation angular powers are almost the same, as shown in the top panel of Figure \ref{cart2}. 
Thus, the observed total spectrum is almost $N_{e,0}$ times of the single spectrum with $\theta=0$, e.g., $F_{\nu}\propto\nu^{2/3}$, see Eq.(\ref{scr}). 
On the other hand, if $\nu\gg\nu_{\rm br}$, e.g., $\sigma_p\gg\theta_c(\nu)$, along the line of sight, one should consider the integral spectrum from electrons with different directions. 
For example, electrons with momenta along the line of sight contribute the most to the radiation, while electrons with momenta deviating from the line of sight emit less radiation, as shown in the bottom panel of Figure \ref{cart2}.
For normally-distributed electrons, the electron number per solid angle in $\sigma_p$ is almost the same. Therefore, the integral process would be approximately equivalent to Eq.(\ref{syn}), finally leading to $F_{\nu}\propto\nu^{1/3}$. 
For $\nu\gg\nu_{\rm cr}$, the spectrum would show the classical exponential decay of synchrotron radiation. 

\begin{figure}
\centering
\includegraphics[angle=0,scale=0.4]{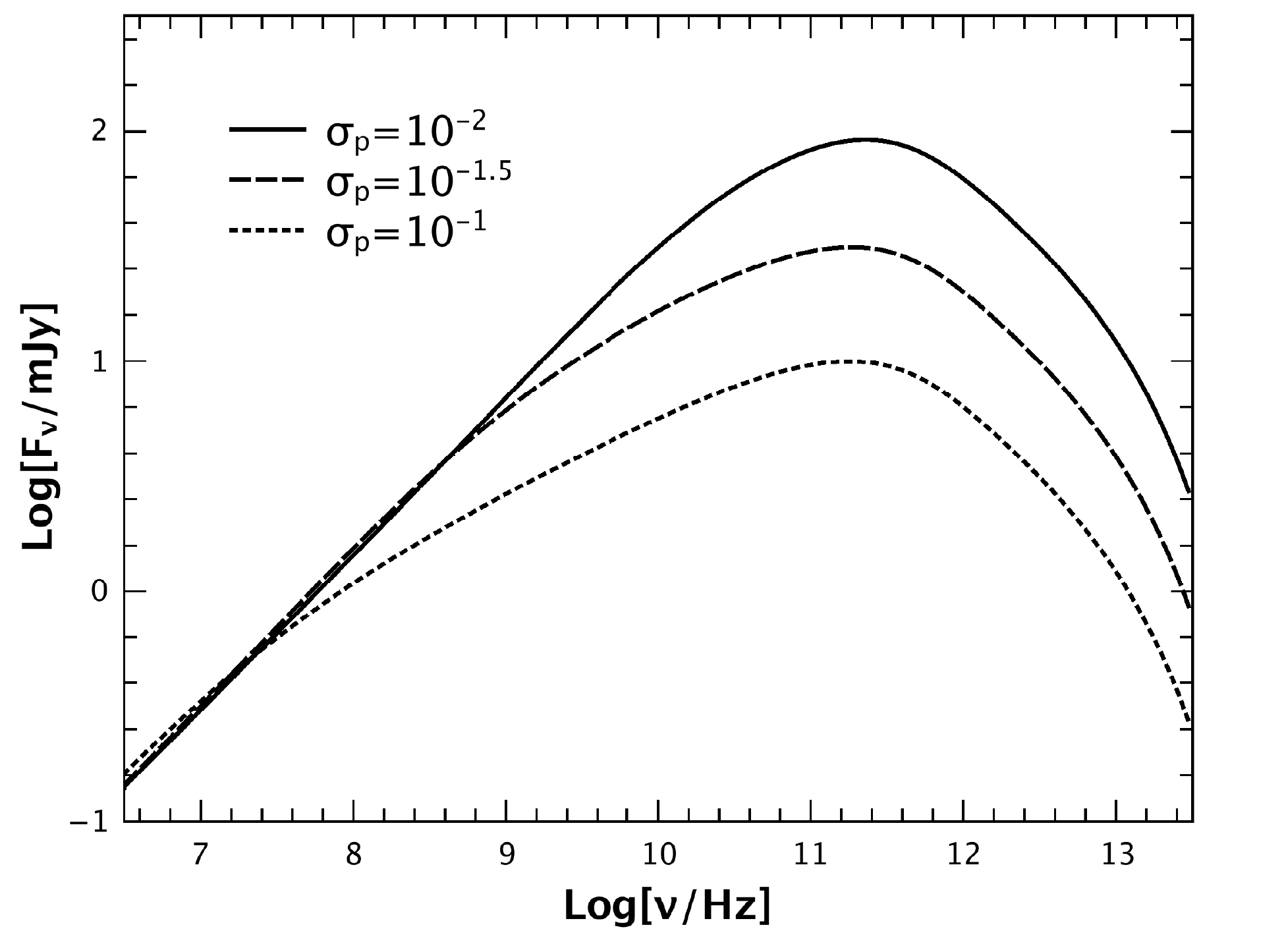}
\includegraphics[angle=0,scale=0.4]{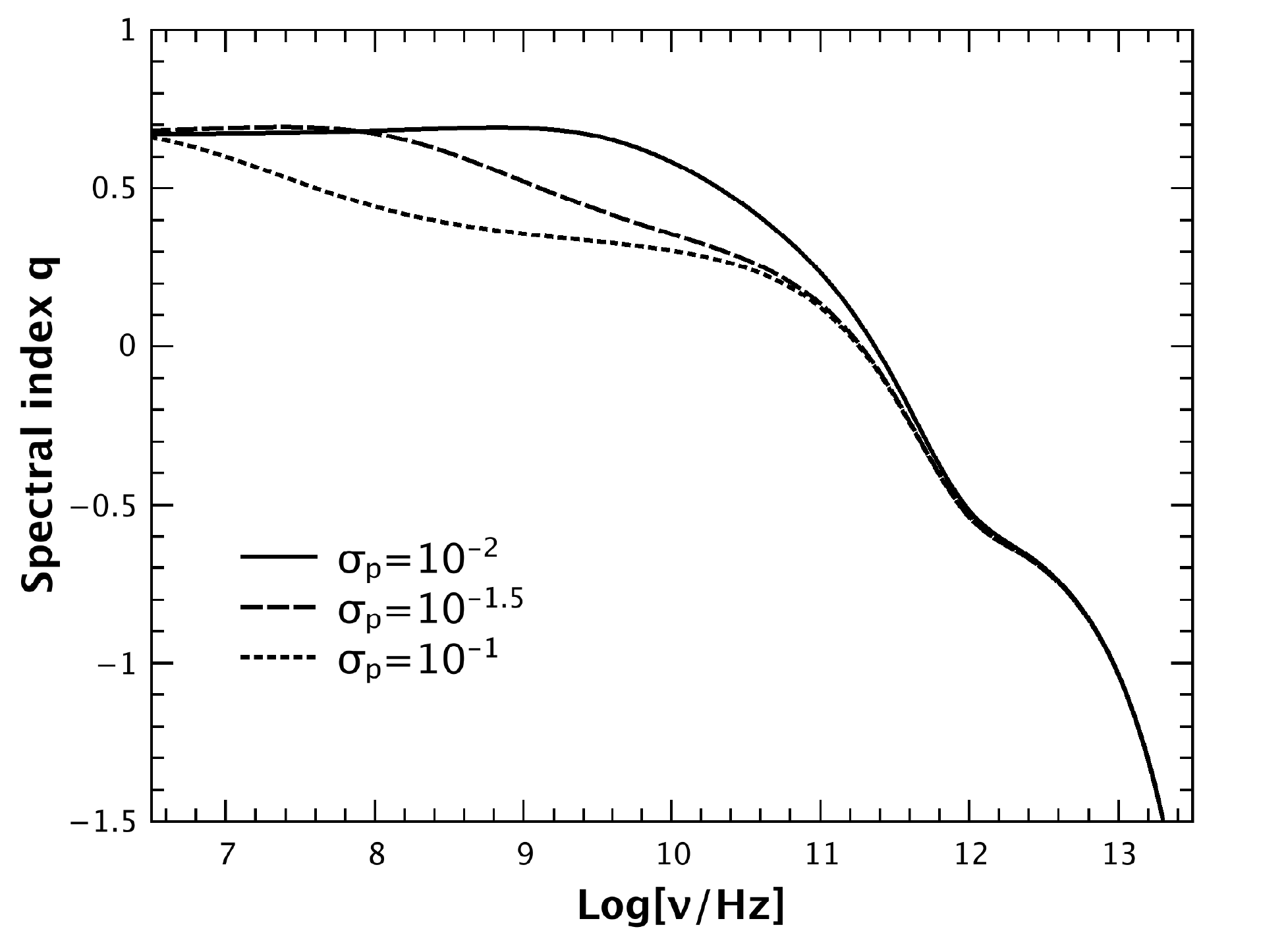}
\caption{Top panel: synchrotron spectra from electrons with a power-law distribution of energy. Bottom panel: the relation between spectral index and frequency. Solid, dashed and dotted curves denote $\sigma_p=10^{-2},~10^{-1.5},~10^{-1}$, respectively. Following parameters are adopted: $\gamma_1=10^2$, $\gamma_2=10^3$, $p=2$, $\alpha_{p,0}=\pi/4$, $B=1~\unit{G}$, $\delta_D=10$, $N_{e,0}=10^{48}$, and $D=1~\unit{Gpc}$}\label{fig3}
\end{figure}

A more realistic case is that the energy distribution of the accelerated electrons is a power-law, e.g., Eq.(\ref{power}).
Define $\nu_{\rm cr,1}$ and $\nu_{\rm cr,2}$ are the critical frequencies of the lowest-energy/highest-energy electrons, respectively.
As shown in Figure \ref{fig3}, for the case with $\nu<\nu_{\rm cr,1}$, the spectrum is contributed by the lowest-energy electrons with $\gamma\sim\gamma_1$, and the resulting spectral index is the same as that of mono-energetic electrons, i.e., $F_{\nu}\propto\nu^{2/3}$ for $\nu\ll\nu_{\rm br}$ and $F_{\nu}\propto\nu^{1/3}$ for $\nu\gg\nu_{\rm br}$. On the other hand, if $\nu_{\rm cr,1}\ll\nu\ll\nu_{\rm cr,2}$, the spectrum would be contributed by all the electrons, so that one has $F_{\nu}\propto\nu^{-(p-1)/2}$. Finally, if $\nu\gg\nu_{\rm cr,2}$, the spectrum is contributed by the highest-energy electrons with $\gamma\sim\gamma_2$, and the spectrum would appear as an exponential decay.

In the above discussion, an important assumption is that the magnetic field is large-scale uniform, which requires that the deflection angle of field lines is less than $1/\gamma$ in the radiation region.

\section{Applications to Blazars and GRBs}

\subsection{Blazars}

Blazars, as the most extreme class of active galactic nuclei, are characterized by a non-thermal continuum spectrum that extends from radio to gamma-rays. The spectral energy distributions of blazars have two broadly peaked components.
The first peak in the IR/UV or even X-ray band is attributed to synchrotron emission by ultra-relativistic electrons in the jet. The second peak, covering X-rays to gamma-rays, is believed to be produced through synchrotron-self-Compton, external Compton, or even hadronic processes \citep[e.g.][]{bot13}.

The radio observations between $74~\unit{MHz}$/$325~\unit{MHz}$ and $1.4~\unit{GHz}$ showed that the values of the spectral index $q$ (defined as $F_\nu\propto\nu^{q}$) are between $-1$ and $1$ for a large fraction of blazars \citep{mas13a,mas13b}.
In a few cases, $q>1/3$ at low-frequency band is observed. According to our theory, this might result from the anisotropic distribution of the electron pitch angles.
For some blazars, the lowest Lorentz factor of electrons in a relativistic blob could be as small as $\gamma_1\sim2-100$ \citep[e.g.][]{kat08,asa18}. If the observed frequency, $\nu_{\rm obs}$, is less than the break frequency $\nu_{\rm br}$ (see Eq.(\ref{break})), e.g., $\sigma_p\lesssim(1/\gamma_1)(2\nu_{\rm cr}/\nu_{\rm obs})^{1/3}\sim0.01-0.5$, the observed spectral index could approach $q\sim2/3$. Such a value of $\sigma_p$ is modestly small. On the other hand, since the minimum Lorentz factor of electrons is small, the condition that the deflection angle of field lines is less than $1/\gamma_1$ could be satisfied.

\subsection{GRBs}

The radiation mechanism of GRBs remains unidentified since they were discovered. The observation showed that the prompt emission spectrum could be well fitted via a Band function \citep{ban00} which appears a smoothly connected broken power ($dN/dE\propto E^{\alpha}$ for $E\ll E_0$ and $\propto E^{\beta}$ for $E\gg E_0$, where $E_0$ is the break energy), and the typical low-energy and high-energy photon spectral indices are $\alpha\sim-1$ and $\beta\sim-2.2$, respectively. Note that here $\alpha$ and $\beta$ correspond to the ``photon'' spectral indices, e.g., $\alpha=q-1$, where $q$ is defined as $F_\nu\propto\nu^q$. 
The classical synchrotron radiation in relativistic shocks is often suggested as a possible mechanism for the prompt emission spectrum \citep{ban00,pre00,nav11,zha11,kum15}. 
In order to explain the observed low-energy photon spectral index $\alpha\sim-1$ (e.g. $q\sim0$) in the majority of GRBs, many attempts have been made, e.g., \citet{bra94,lia97,mes00,pee06,asa09,dai11,uhm14,gen18,xu17,xu18}.

The above models focused on the explanation of $\alpha\sim-1$ for most GRBs under the classical synchrotron theory. However, there are a fraction of GRBs having the spectra with $\alpha>-2/3$ (e.g., $q>1/3$), which has been regarded  not be explained by classical synchrotron theory, so-called ``line-of-death problem'' \citep{pre98,nav11}.
For understanding such a spectral hardness, some modified synchrotron radiation and even other radiation mechanisms, were proposed, including synchrotron self-absorption \citep{pre98}, jitter radiation \citep{med00}, small pitch-angle synchrotron emission \citep{llo00,llo02}, or inverse Compton scattering \citep{lia97}. 

Since electrons with an anisotropic pitch-angle distribution can also generate a spectrum with $\alpha>-2/3$, we consider its application to GRB prompt emission.
The typical observed energy of the prompt emission of GRBs is about a few hundreds keV. In general, for synchrotron radiation, one can write\footnote{For a detailed discussion of synchrotron radiation parameters for GRBs, see \cite{zha02}.}
\be
E_p&\simeq&\Gamma\gamma^2\frac{heB}{2\pi m_ec}\nonumber\\
&\simeq&116~\unit{keV}\left(\frac{\Gamma}{100}\right)\left(\frac{\gamma}{10^4}\right)^2\left(\frac{B}{10^3~\unit{G}}\right),
\ee
where $\Gamma$ is the Lorentz factor of the relativistic shell.
For a relativistic shell with $\Gamma\sim100$ and $B\sim10^3~\unit{G}$, the Lorentz factor of the electrons needs to satisfy $\gamma\sim10^4$. Therefore, if $\sigma_p\lesssim1/\gamma\sim10^{-4}$, the low-energy spectrum could be as hard as $F_\nu\propto\nu^{2/3}$.
However, the condition of $\sigma_p\lesssim1/\gamma\sim10^{-4}$ seems very demanding: 1. the momentum distribution has to be extremely anisotropic when electrons are accelerated; 2. the magnetic field has to be almost parallel to each other so that the deflection angle of field line is much less than $\sim1/\gamma\sim10^{-4}$. If such conditions can be satisfied in the emission region of GRBs, the line of death for synchrotron radiation may be overcome.

In the Internal-Collision-induced MAgnetic Reconnection and Turbulence (ICMART) model \citep{zha11}, the observed emission is the superposition of many mini-jets in the emission region. The lightcurve is the superposition of a broad pulse (defined by the global emission of the ejecta as it streams outwards) and many small-timescale spikes (emission from the mini-jets \citep{zha14}. The broad-pulse is asymmetric with the later part of the decaying segment defined by the curvature effect \citep{zha14,uhm16}. The curvature effect tail extends to the X-ray band, defining a steep-decay phase with rapid spectral evolution consistent with the observations \citep{zha07}. The pitch-angle effect presented above does not modify such a general picture significantly. First, if $\sigma_p > 1/\gamma$ or the field line is too curved, synchrotron radiation would resume the original form with $q=1/3$ so that there is no modification. Second, if the strict condition is satisfied so that $q=2/3$ in each mini-jet is realized, only the slope of very early phase of the steep decay phase is modified (for the steep decay phase with $F_{\nu, {\rm obs}}\propto\nu^{-\hat\beta}t^{-\hat\alpha}$, decay index $\hat\alpha$ is steeper than $2+\hat\beta$, with $\hat\beta=-2/3$ rather than $-1/3$). Due to the very small $\sigma_p$ required by this model, the mini-jets tend to be very spiky. The observed prompt emission spectra may be dominated by the on-beam mini-jets, so that the traditional $q=1/3$ line of death can be overcome. More detailed simulations are needed to verify this.

\section{Conclusions and Discussions}
In this work, we study synchrotron radiation from electrons with a pitch-angle distribution, and apply it to blazars and GRBs. 
Following new conclusions are obtained:
\begin{itemize}
\item {
We consider that the sines of pitch angles of the electrons satisfy a normal distribution with a scatter $\sigma_p$, and the line of sight is along the direction where electron distribution is the maximum. Due to an observational selection effect (bright bursts tend to be detected), such a geometry may be the most relevant geometry for observers. We find that for electrons with a Lorentz factor $\gamma$, if $\sigma_p\lesssim1/\gamma$, the observed flux satisfies $F_\nu\propto\nu^{2/3}$ for $\nu\ll\nu_{\rm cr}$; if $\sigma_p\gg1/\gamma$, the spectrum below $\nu_{\rm cr}$ is a broken power law with a break frequency $\nu_{\rm br}\sim2\nu_{\rm cr}/\sigma_p^3\gamma^3$, e.g., $F_\nu\propto\nu^{2/3}$ for $\nu<\nu_{\rm br}$ and $F_\nu\propto\nu^{1/3}$ for $\nu_{\rm br}<\nu<\nu_{\rm cr}$.
}
\item{
The radiation spectra from astrophysical sources beyond synchrotron death line, i.e., $q>1/3$, can be interpreted by invoking electrons with an extremely anisotropic pith-angle distribution, i.e., $\sigma_p\lesssim1/\gamma_1$, where $\gamma_1$ is the minimum Lorentz factor of the electrons. Such anisotropic pitch angles may be realized by invoking particle acceleration in an ordered magnetic field of a relativistic shock wave or magnetic reconnection region. For example, for a relativistic shock wave with Lorentz factor $\Gamma$, in the shock wave frame, the upstream particles would flow into the shock wave with a Lorentz factor $\Gamma$, and the upstream particles would be confined in an angle of $\lesssim1/\Gamma$. Therefore, if the magnetic field is large-scare uniform (the deflection angle of field line is much less than $\sim1/\gamma_1$), the pitch-angle distribution of particles would be confined within the angle $\lesssim1/\Gamma$. 
}
\item{The theory may be applied to blazars and GRBs. For radio emission of blazars, since the minimum Lorentz factor of electrons could be small, e.g., $\gamma\sim1-10$,
the requirements for the pitch-angle distribution and the deflection of magnetic field lines are modest. It is therefore relatively easy to break the synchrotron death line, and the theory can interpret some blazars whose low-energy spectral index is harder than $q=1/3$. For prompt gamma-ray emission of GRBs, since the minimum Lorentz factor of electrons is large, e.g., $\gamma\sim10^4$, the conditions under which our theory applies is more demanding. In any case, if those conditions are satisfied, GRBs that are slightly beyond the traditional synchrotron line of death can be in principle still accounted for within the framework of synchrotron radiation model (e.g. within the ICMART model). The ultimate line of death is pushed to $q=2/3$.
}

\end{itemize}
 
\acknowledgments 
We thank the anonymous referee for detailed suggestions that have allowed us to improve the presentation of this manuscript.
We thank also Kai Wang for helpful discussions. 
This work is partially supported by Project funded by the Initiative Postdocs Supporting Program (No. BX201600003), the National Basic Research Program (973 Program) of China (No. 2014CB845800), and the China Postdoctoral Science Foundation (No. 2016M600851). Y.-P.Y. is supported by a KIAA-CAS Fellowship.

\end{document}